# Active Amplification of the Terrestrial Albedo to Mitigate Climate Change: An Exploratory Study[*]

Robert M. Hamwey[†]

Cen2eco: Centre for Economic and Ecological Studies, Geneva, Switzerland

**Abstract:** To date, international efforts to mitigate climate change have focussed on reducing emissions of greenhouse gases in the energy, transportation and agriculture sectors, and on sequestering atmospheric carbon dioxide in forests. Here, the potential to complement these efforts by actions to enhance the reflectance of solar insolation by the human settlement and grassland components of the Earth's terrestrial surface is explored. Preliminary estimates derived using a static two dimensional radiative transfer model indicate that such efforts could amplify the overall planetary albedo enough to offset the current global annual average level of radiative forcing caused by anthropogenic greenhouse gases by as much as 30% or 0.76 $Wm^{-2}$. Terrestrial albedo amplification may thus extend, by about 25 years, the time available to advance the development and use of low-emission energy conversion technologies which ultimately remain essential to mitigate long-term climate change. While a scoping analysis indicates the technical feasibility of sufficiently enhancing human settlement and grassland albedos to levels needed to achieve reductions in radiative forcing projected here, additional study is required on two fronts. Firstly, the modelled radiative forcing reductions are static estimates. As they would generate climate feedbacks, more detailed dynamic climate modelling would be needed to confirm the stationary value of the radiative forcing reduction that would result from land surface albedo amplification. Secondly, land surface albedo amplification schemes may have important economic and environmental impacts. Accurate *ex ante* impact assessments would be required to validate global implementation of related measures as a viable mitigation strategy.

## 1. Introduction

The negative environmental and economic impacts of anthropogenic climate change, and the inherent difficulties of reducing causative greenhouse gas accumulations in the atmosphere, are now widely acknowledged. Greenhouse gas emissions – an unavoidable by-product of oil, gas and coal energy conversion – occur principally in the form of carbon dioxide ($CO_2$). Since the onset of industrialisation in 1750, energy-related $CO_2$ emissions have accumulated in the atmosphere, raising the $CO_2$ concentration level from 278 ppmv in 1750 to 378 ppmv at the end of 2004 (Keeling and Whorf 2005). Rising atmospheric concentrations of $CO_2$ and other greenhouse gases increase the radiative forcing of the climate system (Hansen et al. 1997; Ramaswamy et al. 2001) that leads to climate change.

The Intergovernmental Panel on Climate Change (IPCC) has estimated that the total radiative forcing of the climate system due to anthropogenic emissions of long-

---

arXiv:physics/0512170 v1 19 Dec 2005

lived and globally mixed greenhouse gases was 2.43 $Wm^{-2}$ ± 10% in 1998 (Houghton et al. 2001). $CO_2$ emissions resulting from fossil fuel combustion were estimated to account for 60%, or 1.46 $Wm^{-2}$, of this anthropogenic forcing, making it the dominant human-influenced greenhouse gas. More recent studies have indicated that total anthropogenic radiative forcing has risen to a current value of over 2.50 $Wm^{-2}$ while the relative contribution of $CO_2$ to this total has increased to 62% (Hofmann et al. 2005).

In view of their dominant role in precipitating adverse climate changes, international mitigation efforts have largely focussed on reducing emissions of $CO_2$ through energy efficiency and conservation measures, and on sequestering atmospheric $CO_2$ through afforestation programmes (UNFCCC 2003). These activities continue to be motivated by national policies pursuant to the United Nations Framework Convention on Climate Change (UNFCCC), and more recently, by the Convention's Kyoto Protocol. However, analyses suggest that achievement of *current* Kyoto Protocol reduction targets is at best consistent with a trajectory to future stabilisation of $CO_2$ concentrations at 550 ppmv, a level substantially higher than current values and one that falls short of avoiding significant adverse climate change during the current century (Nakicenovic and Swart 2000). This points to the need for stronger mitigation efforts, not only through higher emissions reduction targets and wider international participation in emissions reduction activities, but also through novel approaches to climate change mitigation. Within this context, the IPCC has recognised that, in addition to greenhouse gas emissions reductions, geoengineering offers a potential approach for mitigating changes in the global climate (Apps et al. 2001).

Geoengineering involves large scale and purposeful efforts to circumvent the anthropogenic greenhouse effect by actively managing the energy balance of the Earth (NAS 1992; Flannery 1997; Keith 2000). A common element of many geoengineering schemes involves reducing the incident radiative flux of solar energy in the lower atmosphere and at the Earth's surface to offset the warming effect of greenhouse gases. In a recent study, it was shown that the geoengineering schemes that reduce incident solar radiation uniformly by 1.8% could largely mitigate global and annual mean climate change resulting from a doubling of atmospheric $CO_2$ concentrations from pre-industrial levels (Govindasamy and Caldeira 2000).

To date, geoengineering proposals have typically been of high technological content and cost, implemented at the macro-level by governments, and centred on schemes to enhance terrestrial carbon sinks or partially shield the Earth's surface from sunlight (Flannery et al. 1997; Keith 2001). The latter have included schemes to increase the planetary albedo of the Earth by injecting aerosols into the atmosphere and deploying an array of metallic balloons in the stratosphere or reflective mirrors in Earth orbit.

Terrestrial albedo amplification through land surface modification described here is distinctly different from previous 'geoengineering' proposals in two ways. Mechanistically, it aims to increase the amount of solar radiation reflected by the Earth's surface rather than reducing incident solar radiation flux. Secondly, in practical terms, land surface modification schemes have relatively low technological content and are based on the collective voluntary efforts of local actors rather than a centralised effort of government institutions. These features imply that land-based surface albedo amplification costs could be relatively low and distributed among the participating global population, making the approach attractive from both an economic and implementation perspective.

## 2. Land-based Surface Albedo Amplification as a Mitigation Option

Over the past decade, a number of studies have demonstrated that land use changes resulting from natural and human activities can significantly modify radiative climate forcings on regional scales by changing surface albedos and the energy budget of the lower atmosphere that they regulate (Charney et al. 1977; Bonan et al. 1992; Henderson-Sellers et al. 1993; Xue and Shukla 1993; Myhre and Myhre 2003). Such studies have examined the effects on climate of regionally specific land-use changes. For example, it has been shown that during the past two centuries, regional-scale replacement of natural forests by agricultural crops in the continental United States has increased surface albedos and reduced radiative forcing (Bonan 1997). The climatic effects of projected future land use changes have also been examined. One study has shown that increased radiative forcing arising from decreases in surface albedos associated with large-scale boreal and temperate forestation programs, may offset the climate change mitigation effects of carbon sequestration underpinning such programs (Betts 2000).

In addition to albedo related radiative effects, many of these studies have shown that regional changes in surface albedo also trigger important climatic feedbacks, including changes in regional hydrological cycles resulting from modified evapotranspiration patterns of soil and vegetation. Detailed climate models coupling general atmospheric circulation and land surface models have been advanced to examine the climatic effects of different land use patterns (Bonan 1995). These models capture not only the changes in the surface energy budget arising from changes in surface albedo, but also the associated climate feedbacks resulting from induced changes in latent, sensible and sub-surface heat flux (Bonan 1996).

A major finding emerging from research on the climatic effects of land use patterns is that the regional climate forcing caused by modern land use practices can be comparable to other anthropogenic climate forcings including those resulting from increased greenhouse and aerosol emissions (Bonan 1997). This suggests that there may be significant potential to offset greenhouse gas climate forcing through – as yet unexplored – intentional efforts to increase the Earth's land surface albedo on a global scale by modifying the radiative characteristics of human settlements and vegetation.

## 3. Methods, Model Description and Results

The terrestrial albedo amplification scheme described here is motivated by the observation that just as greenhouse gas emissions that cause climate change have grown with increasing world population, so too have the surface areas of settled land and human-managed grasslands and their contributions to the value of the overall planetary albedo. Approximations of the current magnitude of these contributions are combined with global demographic and vegetation data, and estimates of the extent to which land surface albedos may be increased, to construct globally amplified terrestrial albedo distributions for postulated amplification scenarios. These distributions are then convolved with solar insolation data to model first order spatially resolved radiative forcing reductions achievable under each albedo amplification scenario. In addition, the potential of these radiative forcing reductions to offset the positive radiative forcing associated with rising concentrations of greenhouse gases in the atmosphere is evaluated. Two amplification scenarios are considered. Scenario 1 involves enhancing the albedo of human settlements (manmade surfaces). Scenario 2 involves enhancing the albedo of the world's grasslands.

A first order estimate of the magnitude of a perturbation in the radiative forcing, $\Delta F$, of the Earth's climate due to changes in the shortwave (0.2 – 4.0 µ) surface albedo can be readily determined when the magnitude of the all-sky downward shortwave solar flux at the Earth's surface, $\Phi$, and the perturbed, $\tilde{a}$, and unperturbed, $a$, values of the all-sky shortwave surface albedo are specified for each point and time on the Earth's surface (Hartmann 1994). In this first order approximation, which ignores feedback perturbations arising from changed surface albedos (see Section 4 for a discussion), the decrease in global radiative forcing at an element $s$ of the Earth's surface due to albedo amplification at any time $t$ is given by:

$$\delta F(s,t) = [\tilde{a}(s,t) - a(s,t)] \Phi(s,t), \tag{1}$$

and the annual average decrease in global radiative forcing is:

$$\Delta F = \frac{1}{T} \int_T \int_S \delta F(s,t) ds\, dt = \frac{1}{T} \int_T \int_S [\tilde{a}(s,t) - a(s,t)] \Phi(s,t) ds\, dt, \tag{2}$$

where the perturbation contributions from each surface element, $s$, are integrated over the Earth's total surface $S$ over an entire year $T$. A discrete monthly ($t \in [1,12]$; $T = 12$), radiative transfer model of 1 x 1 degree latitude-longitude resolution was constructed to derive the annual average estimates of $\Delta F$ reported here.

All data used in the model were obtained from the International Satellite Land-Surface Climatology Project (ISLSCP) Initiative II Data Archive (Hall et al 2005). ISLSCP data are derived from various sources (the original source of each dataset is noted below). All datasets have a spatial resolution of 1 x 1 degree, and all model results were obtained from computations covering the 64,800 1 x 1 degree cells defining the gridded 180 degree latitude by 360 degree longitude Earth surface. The NASA EOS Land Mask was used to discriminate between land and water cells.

Monthly average all-sky downward shortwave solar flux at the Earth's surface, $\Phi(s,t)$, and monthly all-sky surface albedos, $a(s,t)$, from the WCRP/GEWEX surface radiation budget (SRB) project release 2 (Stackhouse et al. 2001) were used as inputs in the model. Both of these SRB parameters are measured over the 0.2 – 4.0 µ solar spectrum. Monthly SRB data covering the entire January through December 1986 were used in the model. Surface albedo perturbations were applied only to snow-free cells. The presence of snow in a cell during any given month was determined from monthly snow cover data for 1986 from NASA (Armstrong et al. 2003). Since only SRB all-sky radiation parameters were used in the model, and because the derivation of these data within the SRB project implicitly accounts for the effect of monthly cloud transmittance, there was no need to monitor each cell's monthly configuration of clear-sky and overcast conditions for independent clear-sky and overcast surface radiation flux calculations in the model.

The SRB data were used to compute monthly values of the unperturbed (i.e., baseline state) net radiative shortwave flux at the Earth's surface due to surface absorption of incident shortwave solar radiation for each cell. To simulate active albedo amplification, positive perturbations in the surface albedo of land cells were introduced in the model (as described below). The difference in net radiative flux between the baseline and perturbed surface albedo states was calculated monthly for each cell (using Eq. 1 above) to estimate monthly perturbations in radiative forcing for each cell, and by integrating these, for each 1 degree zonal band from -90 to 90 degrees latitude. From these data, spatially resolved and (surface area weighted) zonal global annual average decreases in radiative forcing were derived.

Surface albedo perturbations introduced in the model were set for each cell based on values of human population and grassland fractions characterising each cell and specifications set in the albedo amplification scenarios on how each cell's surface albedo changes in proportion to these values. Accurate and current spatially resolved population and land cover data thus formed essential inputs to the model.

Global data on population in 1995 from the Gridded Population of the World (GPW) version 2 dataset provided population counts in each land cell over the Earth's surface for a total world population of ~ 5.67 billion in 1995 (Balk et al. 2004, CIESEN 2000). To simulate the current 2005 global population distribution, the 1995 GPW population data where uniformly scaled up by a factor of 1.14; the ratio of estimated 1995 to 2005 global population (USCB 2005). Land cover data from the United States Geological Survey's (USGS) Global Land Cover Characterization (GLCC) classifies the fraction of International Geosphere Biosphere Programme (IGBP) vegetation zones in each cell based on 1 km resolution satellite observations covering the 12-month April 1992 to March 1993 period (Loveland et al. 2000). Both the 1 degree and 1 km resolution data were analysed to confirm consistency, however, only the 1 degree resolution data was used in the model. The current 2005 distribution of world vegetation zones was assumed not to have changed significantly from the 1992-93 distribution used to derive the GLCC dataset.

### *3.1) Scenario 1: Amplification of human settlement albedos*

Under scenario 1, in order to compute the total surface area of human settlements capable of being 'whitened' in an albedo amplification effort, an estimate of the average per capita artificial surface area that makes up human settlements must be made. The area of the terrestrial surface occupied by human infrastructure is not accurately known. Although the resolution of currently characterised global land surface imagery (~1km$^2$) is capable of capturing urban areas, it does not capture dispersed human infrastructure outside of urban agglomerations. A world total of approximately 260,000 km$^2$ of urban and built-up area under the IGBP classification scheme was derived from 1 km GLCC analyses undertaken here, indicating a global average of 46 m$^2$ of urban surface per capita in 1995.

The magnitude of total (urban and dispersed) artificial surface area per capita significantly exceeds the value of urban surface area per capita. Artificial surface area is generally considered to include all residential, recreational, industrial, commercial, transportation-related and institutional land (occupied by man-made physical structures and adjoining landscaped areas) but excludes agricultural land (USDA 2003). Based on this definition, estimates point to a global average value of 440 m2 to 500 m2 of artificial surface per capita (UNEP/RIVM 2004; Wackernagel et al 2002). Moreover, estimated values of artificial surface per capita vary considerably among regions. The highest regional estimate of artificial surface per capita is for United States where it is estimated to exceed 1,500 m$^2$ per capita (USDA 2003), whereas for East and South Asia, regional estimates are the lowest at about 300 m$^2$ per capita (UNEP/RIVM 2004). Based on these estimates, a global average per capita reflective surface area for human settlements, $\sigma$, of 500 m$^2$ was adopted in this study.

A composite surface albedo for human settlements, $a_h$ = 0.15, a typical urban value (Jin et al. 2005; Taha 2005), was assumed globally as the unperturbed surface albedo of human settlements. To simulate the perturbed state, the surface albedo of human settlements was enhanced globally by 100% to a higher value of $\tilde{a}_h$ = 0.3, resulting from an intentional 'whitening' of human structures as a climate change mitigation measure. In the model, $\sigma$, $a_h$ and $\tilde{a}_h$ were assumed to be invariant with respect to solar zenith angle and season.

Using the adopted values of $\sigma$, $a_h$ and $\tilde{a}_h$, the surface albedo values of cells were revised upward in proportion to the human population in each cell. For each cell $s$ and month $t$ the perturbed surface albedo was computed as:

$$\tilde{a}(s,t) = SM(s,t) \times \left[\frac{\sigma P(s)}{SA(s)}\right] \times (\tilde{a}_h - a_h) + a(s,t), \qquad (3)$$

where $SM(s,t)$ is a no-snow mask equal to 1 (0) if snow is absent (present) in cell $s$ during month $t$; $P(s)$ is the estimated 2005 population, and $SA(s)$ is the surface area, of cell $s$; and $a(s,t)$ is the baseline surface albedo of cell $s$ during month $t$. The decrease in radiative forcing in each cell $s$ during month $t$ immediately follows from Eq. 1.

Monthly changes in surface albedo and radiative forcing were calculated for each cell. The baseline data indicate an annual average global all-sky surface albedo of 0.140804. In response to the increase in the surface albedos of human settlements under scenario 1, the model projects this figure to increase by 0.000875, and the annual globally averaged radiative forcing to decrease by 0.17 Wm$^{-2}$.

*3.2) Scenario 2: Amplification of grassland albedos*

Under scenario 2, amplification of the planetary albedo is achieved by increasing the surface albedo of the world's grasslands. The 'grassland' area assumed in the model, and referred to hereafter simply as grasslands, includes 3 IGBP classifications: open shrubland, grasslands and savannah. Taken together, about 30% of the Earth's land area falls under these three IGBP categories.

It is not possible to define a common baseline value of unperturbed grassland surface albedo for all model cells due to latitudinal and seasonal variations in this quantity that result from temporal variations in cloud cover, mean solar zenith angles, precipitation and phase offsets of annual vegetation growth cycles. Therefore, an algorithm was used to dynamically specify grassland baseline surface albedos. Within the model, the mean monthly values of surface albedo for cells in each 1 degree latitude zonal band with a grassland fraction of 80% or more were calculated. These values were used to define the baseline surface albedo of grasslands as a function of latitude and month. For a minority of latitude bands without cells containing a grassland fraction of 80% or more, the baseline value was set to that of the nearest latitude band in which this condition was met. Through this procedure, a set of unperturbed baseline surface albedo values of cell grasslands, $a_g(s,t)$, was established. The global annual average all-sky grassland surface albedo derived from the SRB dataset was 0.17.

Under scenario 2, the baseline surface albedo of grasslands, $a_g(s,t)$, was increased by 25% to a higher value of $\tilde{a}_g(s,t)$. The perturbed surface albedo for each cell $s$ and month $t$ was computed as:

$$\tilde{a}(s,t) = SM(s,t) \times \left[\frac{GF(s)}{SA(s)}\right] \times (\tilde{a}_g(s,t) - a_g(s,t)) + a(s,t), \qquad (4)$$

where $GF(s)$ is the grassland fraction of the cell $s$. The decrease in radiative forcing in each cell $s$ during month $t$ immediately follows from Eq. 1.

Monthly changes in surface albedo and radiative forcing were calculated for each cell. In response to the increase in the surface albedos of grasslands under scenario 2, the model projects an increase in the annual average global all-sky surface albedo of 0.002626 and a decrease in annual globally averaged radiative forcing of 0.59 Wm$^{-2}$. Such a large decrease is not unexpected considering that 30% of the Earth's land surface is occupied by grasslands, particularly at low and middle latitudes where the value of annual insolation is highest. However, because human intervention to enhance the surface albedo of grasslands may be feasible for only the portion of the world's grasslands that can be managed by man, the 0.59 Wm$^{-2}$ decrease in radiative forcing achievable under scenario 2 represents an upper limit for a presumed uniform global grassland surface albedo increase of 25%. If albedo enhancement is feasible only for a fraction, $\beta$, of all grasslands, and this subset has the same spatial distribution as the larger set of world grasslands, scenario 2 would result in a decrease in annual globally averaged radiative forcing of 0.59$\beta$ Wm$^{-2}$.

### *3.3) Detailed results*

Monthly spatial and latitudinal estimates of decreases in radiative forcing were computed by the model for scenarios 1 and 2. Spatial maps of the annual average decrease in radiative forcing resulting from scenario 1, scenario 2, and scenario 1 and 2 combined, are presented in Figure 1. The annual average decreases in radiative forcing resulting from scenario 1 and 2, as a function of latitude, are presented in Figure 2. These results are discussed further below.

## 4. Qualification of Results: Climate Feedbacks

The radiative transfer model used in this study only provides an estimate of changes in the shortwave surface radiation budget. The full surface energy budget includes other parameters that are external to the model. Specifically, under steady-state conditions, the full surface energy balance equation for a cell *s* is:

$$S_{net} + L_{net} = LH + SH + HF, \qquad (5)$$

where $S_{net} = S\downarrow - S\uparrow$ is net (downward – upward) shortwave radiation flux, $L_{net} = L\downarrow - L\uparrow$ is the net (downward – upward) longwave radiation flux, LH and SH are respectively the latent and sensible heat fluxes from the surface to the atmosphere, and HF is the horizontal sub-surface flux of heat absorbed at the surface to adjacent cells. To maintain the energy balance in Eq. 5, the radiative forcing perturbations in cells generated under scenarios 1 and 2, $\delta F = \delta S_{net} < 0$ will result in climatic feedbacks:

$$\delta F = -\delta L_{net} + \delta LH + \delta SH + \delta HF. \qquad (6)$$

In response to the shortwave radiative forcing perturbation arising from increased surface albedos, Eq. 6 indicates that several parameters will adjust in order to maintain radiative equilibrium at the land surface. However, the radiative transfer model used in this study provides no information on how these adjustments are partitioned; a coupled climate land surface model is required to provide such an indication (Bonan 1997).

Nevertheless, assuming that $|\delta L_{net}| \ll |\delta S_{net}|$ and $\delta HF \ll \delta LH \sim \delta SH$, Eq. 6 indicates that evapotranspiration (*LH*) and atmospheric convection (*SH*), and consequently cloud cover and precipitation, may be reduced by increases in regional surface albedos. Such reductions have been examined in previous studies of tropical regions (Charney et al. 1977; Henderson-Sellers and Gornitz 1984; Xue and Shukla 1993; Dirmeyer and Shukla 1994). At the same time, these primary feedbacks will themselves induce secondary feedbacks. If reductions in annual average cloud cover relative to the baseline state are substantial, they may raise absorbed shortwave flux, and reduce absorbed longwave flux due to reduced cloud forcing. This could potentially result in a net increase of annual (shortwave + longwave) insolation and an overall heating effect (Charney et al. 1977), or no significant decrease in temperature (Henderson-Sellers and Gornitz 1984), even though surface albedo has increased.

Whereas the above feedbacks may be important in tropical regions, modelling of modern surface albedo changes of order ~ 0.02 – 0.06 occurring in the temperate United States indicate no significant change in seasonally averaged precipitation and a net cooling effect (Bonan 1997). Moreover, a recent global simulation of regional climate responses to land conversion induced surface albedo increases (DeFries et al. 2002) has highlighted that effects are quite different in the tropics, where net warming and drier hydrological conditions are expected, than in temperate regions where net cooling and largely unaltered hydrological conditions result.

The complex nature of the multiple feedbacks above emphasises the need for accurate coupled general atmospheric circulation and land surface modelling of global surface albedo amplification schemes to examine how long-term surface radiation budgets, energy budgets, and climatologies may ultimately be affected in different regions.

## 5. Potential Mitigation Benefits

The radiative forcing perturbation estimates obtained here suggest that, implemented together, the two albedo amplification scenarios described above could potentially offset 0.76 Wm$^{-2}$, or about 30% of the approximately 2.50 Wm$^{-2}$ of radiative forcing caused by anthropogenic emissions of all long-lived and globally mixed greenhouse gases. However, as Figure 1 shows, decreases in radiative forcing vary considerably by region. Expectedly, the spatial and latitudinal distributions of radiative forcing perturbations under scenario 1 and 2 closely trace the underlying physical distribution of population and grasslands. But as these two distributions are dissimilar, the combined effects of scenarios 1 and 2 result in a distribution of radiative forcing perturbations that extends over a fairly large fraction of the terrestrial surface. Nevertheless, perturbations significantly higher than the global average occur in highly populated and high grassland fraction cells.

As Figure 2 shows, the latitudinal distributions of annual average radiative forcing reductions achieved under scenarios 1 and 2 are mostly concentrated in the Northern Hemisphere. Unlike greenhouse gas emissions which result in nearly uniform latitudinal increases in radiative forcing due to mixing in the atmosphere, the decreases in radiative forcing induced by land surface albedo amplification vary in proportion to the underlying latitudinal distribution of population and grasslands. Projections of the effects of latitude dependent albedo amplification on climate and climate change require study in a general circulation model. Yet a recent study of this type (Govindasamy and Caldeira 2000) suggests that it is not necessary for the latitudinal pattern of radiative forcing due to albedo amplification to match that of greenhouse gases to largely negate the effects of the latter.

The mitigation power of albedo amplification can be compared with that of $CO_2$ emissions reductions by evaluating the decrease in atmospheric $CO_2$ concentration needed to achieve a similar reduction in radiative forcing. A simplified expression for the calculation of radiative forcing induced by $CO_2$ has been validated by the IPCC (Ramaswamy et al. 2001) as:

$$\Delta F = \alpha \ln(C/C_0),  \qquad (7)$$

where $\alpha = 5.35$ and $C$ and $C_0$ are the perturbed and unperturbed $CO_2$ concentrations. This expression indicates that the estimated decreases in radiative forcing of 0.17 and 0.59 $Wm^{-2}$ under scenarios 1 and 2 respectively, are equivalent to decreases obtained by reductions in atmospheric $CO_2$ concentration of 12 and 40 ppmv, or 50 ppmv when the two scenarios are implemented together (note: Eq. 7 is not linear).

Recent observations in both the Northern and Southern Hemisphere indicating a 2 ppmv annual growth rate of atmospheric $CO_2$ concentrations (NIWA 2004; Keeling and Whorf 2005) would, *ceteris paribus*, suggest that the albedo amplification measures modelled here could offset as much as 25 years of global $CO_2$ emissions at current levels (~ 6 yrs under scenario 1 and ~ 20 yrs under scenario 2). Although such an offset would not obviate the need for long term emission reductions to reduce future climate change, it could substantially extend the time available to advance the development and use of low-emission energy conversion technologies. Land surface albedo amplification efforts may thus represent potential options to complement mitigation activities focussing on emissions reductions.

## 6. Implications and Feasibility

### 6.1) Scenario 1: Amplification of human settlement albedos

While the estimates reported here indicate that amplification of human settlement and grassland albedos have the potential to significantly offset radiative forcing caused by anthropogenic greenhouse gases, the scope for, and feasibility of engineering such amplifications on a global scale for climate change mitigation remains largely unexplored. Technologies are readily available to amplify the surface reflectance of human settlements, and these have been the subject of considerable research over the past decade (Rosenfeld et al. 1997) within the context of reducing ambient temperatures and associated energy costs for cooling, and ground-level ozone concentrations, in urban heat islands (UHI). However, the potential of substantially enhancing the albedo of vegetation, particularly grass species, remains unexplored.

Major foci of UHI mitigation have been to increase the albedo of building roofs and facades using high reflectivity titanium dioxide ($TiO_2$) paints and films, and roads and other paved surfaces using high albedo *white cement* (HARC 2004; Taha 2005). Typically, these technologies raise roof albedos from 0.1 to 0.7, and paved surface albedos from 0.1 to 0.4. Conversion costs range from 15-30 $m^{-2}$ for roofs and 15-25 $m^{-2}$ for pavements, while the lifecycle of converted surfaces are on the order of 10 years (HARC 2004). Detailed studies of several urban areas in the United States (Rose et al. 2003) have demonstrated, that widely implemented, these measures are able to increase baseline urban albedos, characterised by values ranging from 0.12 – 0.16, by 100% or more, to values ranging from 0.20 to 0.37, depending on urban land cover category (Taha 2005). The 100% enhancement of human settlement surface albedos postulated in scenario 1 thus appears technologically feasible. Moreover, human settlement surface albedos could be enhanced still further if high albedo grasses are used to increase the albedo of lawn surfaces (see below).

Studies on albedo enhancement abatement costs (i.e., as a climate change mitigation measure in costs per tonne of $CO_2$ equivalent avoided) would need to be completed, and estimated costs compared with other abatement options to gauge the economic efficiency of any globally implemented measure to enhance the albedo of human settlements. Albedo enhancement abatements costs should internalise the economic co-benefits of energy savings and reduced ground level ozone concentrations that surface albedo enhancements generate in urban heat islands.

Measures to increase the reflectance of human settlements may benefit from high levels of public acceptability since, unlike energy conservation measures, they incur only periodic fixed costs and do not reduce energy consumption utility. Moreover, with increasing public awareness of climate change, the perceptible nature of albedo enhancement activities could encourage widespread community engagement by offering local actors a visible way to demonstrate their contribution to climate change mitigation. Experience with UHI mitigation activities in the United States, where they are most advanced, has been marked by high levels of public support.

### *6.2) Scenario 2: Amplification of grassland albedos*

The substantial radiative forcing offset yielded by enhancing the albedo of the world's grasslands could make such measures particularly attractive. However, one of the principal factors limiting the magnitude of plant albedos, including for grasses, is strong absorption of visible light in the solar spectrum by chlorophyll in plant leaves. Yet the possibility of increasing grassland surface albedos with grasses having high reflectances in the visible bandpass (0.4 – 0.8 μ) remains considerable. Since the reflectance of near infrared solar radiation (0.8 – 1.4 μ) by plants is relatively high, typically ranging from 0.4 – 0.7 (Asner et al. 1998; Asner 1998; Sims and Gamon 2002), and because the intensity of terrestrial solar radiation drops off substantially in the far infrared ( > 1.4 μ) where reflectance by plants is lower, grasslands populated by natural or *bioengineered* grasses having increased visible band reflectivities could generate significantly enhanced grassland surface albedos.

There exist a wide variety of naturally occurring light-coloured shrubs and grasses that exhibit high reflectivities over the visible spectrum. The white, light-green and light-yellow colouration of such plants – i.e., increased reflectance of visible light relative to visibly green plants – occurs through: 1) low concentrations of chlorophyll and other pigments in plants' leaves and stems (Sims and Gamon 2002) and/or 2) the presence of trichomes and waxes (that efficiently reflect visible light) on plants' leaf surfaces (Bondada and Oosterhuis 2000; Grant et al. 2003). Leaf thickness also plays a role in visible reflectance (Knapp and Carter 1998).

Among the first class of plants are grasses such as *Carex hachijoensis* and *Chlorophytum comosum*, and shrubs such as *Alpinia zerumbet, Euonymus europaeus,* and *Ficus aspera.* These plants have variegated leaf colours: patches or stripes of their leaves are light-yellow or white in colour (where chlorophyll and other pigments are absent or in low concentrations) and green elsewhere. Up to 60% of their leaf surfaces are light-coloured. Although uncommon in grassland grasses, variegated leaf colours have been observed in *Stenotaphrum secundatum* (Sud and Dengler 2000).

The second class of plants includes 'white-coloured' shrubs such as *Cerastium biebersteinii* and *Senecio cineraria.* Dense trichomes on these plants' surfaces reflect a significant fraction of visible light and thus limit the amount that traverses the trichome layer to become absorbed by chlorophyll in the underlying leaf. Trichome-rich plants are generally robust with a demonstrated potential to adapt to a wide range of geo-climatic conditions.

In addition to exploring grassland albedo enhancement using natural species, the potential of bioengineering high albedo grass (and shrub) species with lowered leaf pigmentation densities and highly reflective leaf surface trichomes and waxes could be explored. However, as discussed below, leaf reflectance is but one of many factors affecting the surface albedo of grasslands.

Theoretical and observational studies of the reflective properties and surface albedo of vegetation, including of grasslands, continue to be pursued extensively by the remote sensing community in order to derive the surface albedo of land surface vegetation – required in climate models – from satellite based spectral observations (Liang 2000). These studies are, however, limited to dominant, naturally occurring vegetation species that are prominent in satellite observations. Detailed spectral and albedo measurements of minor species are sparse in the literature, and simulated spectral and albedo properties of prospective bioengineered vegetation species, such as high albedo grasses, are entirely absent.

The surface albedo of grasslands are influenced by a large number of factors, including: leaf reflectance, leaf morphology, leaf area and density, leaf canopy geometry and surface roughness, and the reflective properties of exposed underlying soil. Each of these grass features provides a potential entry-point for research efforts to enhance the surface albedo of grasslands.

To investigate how each of the above factors influences the surface albedo of grasslands, it is instructive to revisit the definition of surface albedo. Specifically, the surface albedo, $a(s,t)$, at a vegetation element $s$ of the Earth's surface at time $t$ is the instantaneous ratio of the total reflected to total incident solar shortwave radiation flux at the surface. A determination of surface albedo thus involves an integration of all sources of incident flux (from the sun and sky) and outgoing reflected flux (towards the sky) over the entire skyward hemisphere centred on $s$. Surface albedo is thus not an invariant property of the surface because it depends on the angular position of the sun, as well as the reflective and lifecycle-dependent geometrical properties of the vegetation surface, both of which vary temporally.

While the surface albedos of grasslands are infinitely difficult to compute analytically, they are easy to measure as a scalar quantity through field experiments using goniometers. Moreover, angle dependent spectrophotometric observations provide information on the angular distribution of reflectance revealing the bidirectional reflectance distribution function (BRDF) of a grassland sample (Sandmeier et al 1998). The BRDF of an object, which depends on both illumination and viewing angles, is the ratio of reflected to incident radiation for a given wavelength $\lambda$. Its functional form is:

$$BRDF(\lambda, \theta_i, \phi_i, \theta_r, \phi_r), \qquad (8)$$

where $\theta$ and $\phi$ specify the directions of incident (*i*) and reflected (*r*) flux in spherical coordinates. With knowledge of the BRDF, given any all-sky irradiance distribution, the surface albedo can be determined by convolving irradiance with the BRDF and integrating the over the skyward hemisphere.

Returning to scenario 2 of this study, it is of interest to postulate how grassland surface albedos might be enhanced to ascertain whether the adopted albedo enhancement factor of 25% is realistic and feasible. Of the various grassland characteristics that regulate surface albedo, here, the effect of changing only the physical reflectance properties of grass leaves by lowering chlorophyll concentrations is explored to derive an estimate of the potential of such a change to raise grassland surface albedo.

The BRDF contains all of the spectral and geometric reflectance properties of a surface element of grassland. Here it is postulated, that in a first approximation, these functional dependences are independent so that the grassland BRDF is separable and can be written as:

$$BRDF(\lambda,\theta_i,\phi_i,\theta_r,\phi_r) = F(\lambda) \times G(\theta_i,\phi_i,\theta_r,\phi_r), \tag{9}$$

where $F$ is a linear function of the solar reflectance spectrum of a representative grass leaf in the grassland canopy (note that if the BRDF was not separable, the perceived 'colour' of grass would change with viewing angle). BRDF separability is consistent with recent high resolution remote sensing observations (Barducci et al. 2005). The BRDF separability approximation is needed in order to simulate the effect of enhancing only the leaf reflectance properties of grassland grasses, while leaving all of grassland reflectance properties associated with geometrical features (i.e., leaf area and thickness, and canopy geometry and surface roughness) unaffected.

If only the 'colour' of grass is changed, the angular BRDF component of high reflectivity grasses comprising a grassland sample, $G(\theta_i,\phi_i,\theta_r,\phi_r)$, remains unchanged from its baseline state. Accordingly, the ratio of baseline to enhanced grassland surface albedo for any grassland surface element in scenario 1 is given simply the ratio of the integrated shortwave solar reflectance of the high albedo and baseline grasses:

$$\frac{\tilde{a}_g(s,t)}{a_g(s,t)} = \frac{\int_{SW} \tilde{R}(\lambda) I_{sun}(\lambda,s,t) d\lambda}{\int_{SW} R(\lambda) I_{sun}(\lambda,s,t) d\lambda} \tag{10}$$

where $R(\lambda)$ is leaf reflectance and $I_{sun}(\lambda,s,t) = \Phi(\lambda,s,t)$ is the all-sky (i.e., direct and circumsolar) solar irradiance at surface element $s$ during month $t$ and the integrals are taken over the shortwave spectrum $\lambda \in [0.2 - 4.0\mu]$.

Since an extensive search of numerous spectral libraries revealed no published reflectance spectra for shrubs or grass with variegated leaf colours, a synthetic high albedo grass spectrum was constructed. Using data from the USGS Digital Spectral Library (Clarke et al. 2003), the synthetic reflectance spectrum was constructed as a composite of the reflectance spectra of a typical grassland species, *Poa pratensis* (40% contribution), and a white *Petunia* petal (60% contribution). Because the *Petunia* petal is among the most highly reflective plant surfaces, the synthetic reflectance spectrum thus derived may be considered as an upper limit to the reflectance potential of variegated leaf colour high albedo grass.

The reflectance spectra of the baseline grass (*Poa pratensis*) and high albedo grass (synthetic composite) are displayed in Figure 3. Both were convolved against a standard all-sky ground level solar irradiance spectrum (ASTM 2003) in order to estimate the enhancement of surface albedo generated by a synthetic high albedo grass using Eq. 10. An enhancement of 46% was indicated; a value significantly higher than the 25% surface albedo increase postulated under scenario 2. It may be possible that non-synthetic, naturally occurring grasses with variegated leaf colours could also result in albedo enhancements of 25% or more. It is thus considered, that a shortwave radiative forcing reduction of 0.59 $Wm^{-2}$ would be technically achievable for a conversion of the world's grasslands to a high albedo status. At the same time, the portion of world grasslands amenable to conversion would be limited by resource and access constraints, and so a realistic reduction in radiative forcing due to enhanced grassland albedos would likely be less than the 0.59 $Wm^{-2}$ calculated here.

It should be emphasised that should suitable naturally occurring or synthetic high albedo grasses be identified, consideration of their global-scale introduction in grasslands for climate change mitigation purposes would need to include an appraisal of the economic costs of global grassland conversion and maintenance. It remains to be demonstrated that implementation costs per equivalent $CO_2$ emissions reduction are competitive with other mitigation options. Moreover, there would be political implications for grassland albedo enhancement. As many of the world's grasslands are in developing countries, international financing and technology transfer mechanisms to facilitate their implementation of grassland conversions would be needed to defray the likely high costs arising from associated biotechnology intellectual property rights. The UNFCCC Clean Development Mechanism (CDM) may provide such a mechanism.

At a more fundamental level, assessments would need to be made of the *in-situ* impacts of a large-scale introduction of high albedo grasses on regional climates and ecosystems. As discussed in Section 4, their large-scale presence could have appreciable regional climatic effects by modifying *ex ante* configurations of radiative equilibria, evapotranspiration characteristics, convective flow patterns of regional ecosystems, and resulting patterns of regional cloud cover and precipitation. Reduced photosynthetic activity due to lower chlorophyll content of high albedo grasses, and hence decreased carbon uptake, may also modify regional carbon cycles slightly. Although beyond the scope of this preliminary inquiry, a general circulation model (GCM) or regional climate model (RCM) would be required to examine these effects.

In terms of ecological effects, modified grasses could perturb plant-plant and plant-micro-organism interactions in regional ecosystems, potentially altering baseline population densities and geographical boundaries of other plant species in vegetative systems. Plant-animal interactions may also be perturbed, modifying predator-prey, symbiotic and parasitic relationships, as well as the life-cycle and migration patterns, of insects and animals. A complete picture of potential ecosystem effects would require assessments of land surface albedo amplification impacts on biological dynamics to follow up on RCM results.

## 7. Conclusions

A static two dimensional radiative transfer model used in this exploratory study suggests that active amplification of land surface albedo may represent a potential strategy for climate change mitigation that can complement efforts to reduce greenhouse gas emissions. Preliminary estimates that such efforts could amplify the overall planetary albedo enough to offset the current global annual average level of radiative forcing caused by anthropogenic greenhouse gases by as much as 30% or 0.76 $Wm^{-2}$. Terrestrial albedo amplification may thus extend, by about 25 years, the time available to advance the development and use of low-emission energy conversion technologies which ultimately remain essential to mitigate long-term climate change. However, these results are preliminary estimates, serving only to roughly gauge the climate change mitigation potential of active efforts to modify the terrestrial albedo. Climate feedbacks and feedback loops have not been analysed.

More detailed simulations are needed to validate the estimates derived in this study, particularly as improved models (Dai et al. 2003; Tian et al. 2004) and accurate high resolution surface albedo data (Justice et al. 1998; Schaaf et al. 2002) are now becoming available. Higher resolution coupled modelling using GCM and RCM methodologies and accurately measured location-specific urban and grassland surface albedo values, remains desirable. Accounting for longwave and heat flux feedbacks, such modelling would provide improved estimates of the stationary radiative forcing reduction that can achieved by active amplification of the terrestrial albedo. It would also reveal the dynamic meteorological effects of albedo amplification on regional climates.

Based on improved modelling results, assessments of the climatic and ecological impacts, and related socio-economic impacts, of a large-scale introduction of high albedo human settlements and grasslands could be undertaken. If additional research validates global land-surface albedo amplification as a viable, effective and complementary mitigation strategy, future climate change mitigation efforts could be strengthened.

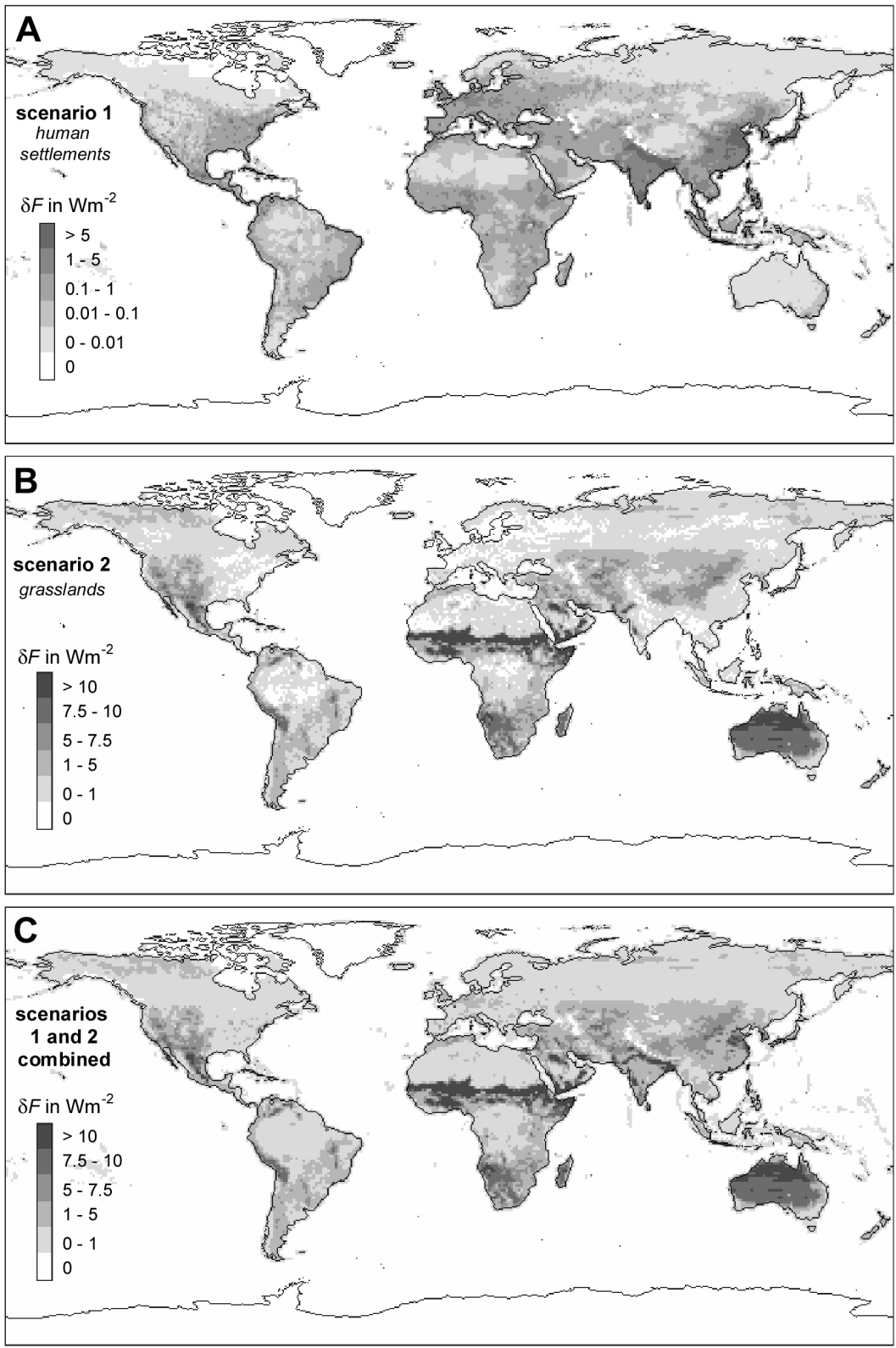

*Figure 1.* Spatial maps of projected decrease in annual average radiative forcing ($\delta F$) resulting from albedo amplification under: (A) scenario 1; (B) scenario 2; and (C) scenarios 1 and 2 combined.

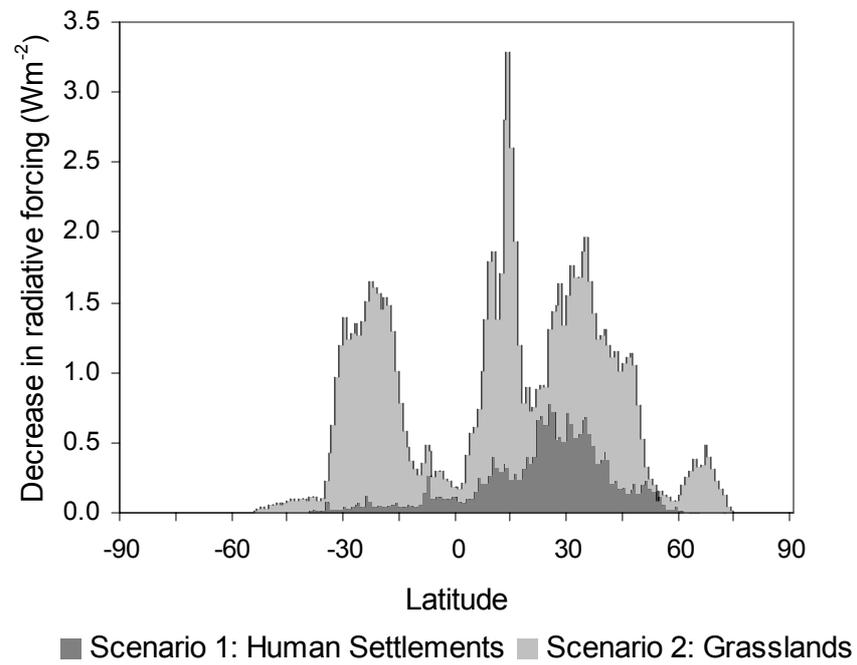

*Figure 2.* Projected decrease in annual average radiative forcing resulting from scenarios 1 and 2 as a function of latitude (stacked histogram).

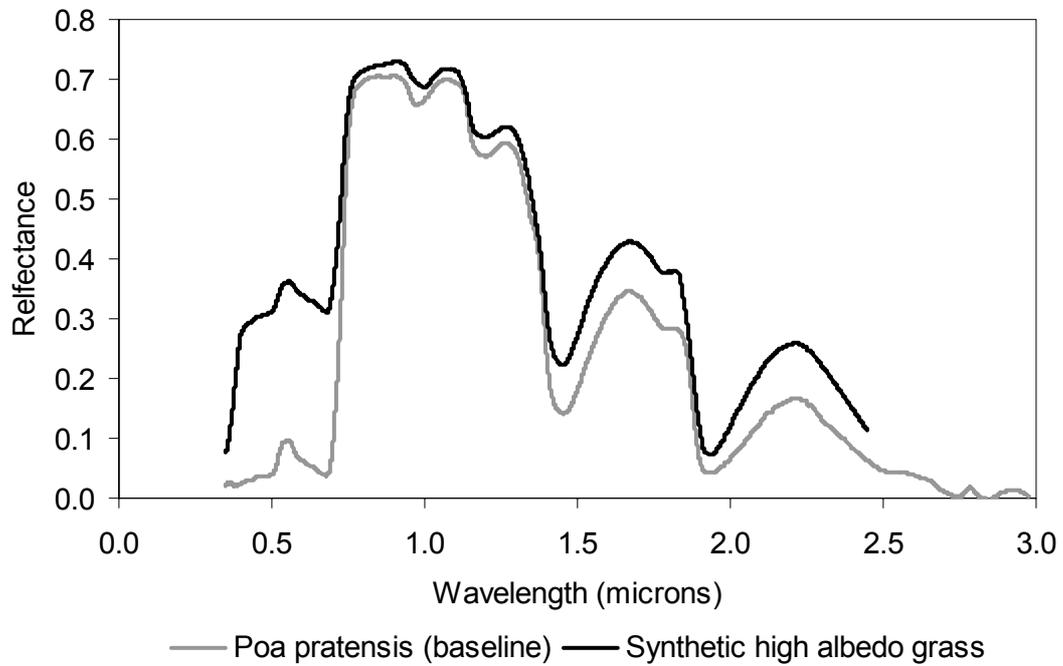

*Figure 3.* Leaf reflectance spectra of baseline and synthetic high albedo grass (see text).